# Melting and Freezing Lines for a Mixture of Charged Colloidal Spheres with Spindle-Type Phase Diagram

Nina J. Lorenz* and Thomas Palberg

Institut für Physik, Johannes Gutenberg Universität Mainz, Staudinger Weg 7, D-55128 Mainz, Germany

Abstract

We have measured the phase behavior of a binary mixture of like-charged colloidal spheres with a size ratio of $\Gamma = 0.9$ and a charge ratio of $\Lambda = 0.96$ as a function of particle number density $n$ and composition $p$. Under exhaustively deionized conditions the aqueous suspension forms solid solutions of body centered cubic structure for all compositions. The freezing and melting lines as a function of composition show opposite behavior and open a wide, spindle shaped coexistence region. Lacking more sophisticated treatments, we model the interaction in our mixtures as an effective one-component pair energy accounting for number weighted effective charge and screening constant. Using this description, we find that within experimental error the location of the experimental melting points meets the range of melting points predicted for monodisperse, one component Yukawa systems made in several theoretical approaches. We further discuss that a detailed understanding of the exact phase diagram shape including the composition dependent width of the coexistence region will need an extended theoretical treatment.

1. Introduction

Charged colloidal spheres in aqueous suspension may undergo a fluid-crystalline phase transition for sufficiently large and long-ranged potentials. Their phase behavior was extensively studied experimentally as the screened Coulomb interactions are conveniently controlled through the choice of the particle radius $a$ and the effective charge $Z^{eff}$ and may be further varied by adjusting the particle density $n$ and concentration of added electrolyte $c_B$ [1, 2, 3]. With increasing $n$ the structure changes from initially fluid to body-centered cubic (bcc) - and finally to face-centered cubic (fcc) crystals. The location of phase transition lines was also predicted theoretically. The phase behavior of Yukawa particles was first obtained by Robbins, Kremer and Grest (RKG) from extensive computer simulations [4]. Later studies by Meijer and Frenkel [5], Stevens and Robbins [6], Du Pont et al. [7], Hamaguchi et al. [8] as well as Hoy and Robbins [9] complemented and extended this work. Experimental results have been compared to the predictions and in general a satisfying agreement could be obtained concerning the first order freezing transition [3, 10, 11]. Interesting open questions concern the location of the bcc-fcc transition and the fluid-bcc-fcc triple point [1, 2, 12] and the width of the coexistence region in the colloid concentration – salt concentration plane of the phase diagram [13, 14].

Recently, the research focus has advanced to mixtures of colloidal spheres, where additional parameters determine the phase behavior. For hard sphere mixtures the size ratio of the small (S) and larger (L) component $\Gamma = a_S / a_L$ and the composition $p = n_S / (n_S + n_L)$, denoting the number fraction of small particles, become important. In charged mixtures, also the charge ratio $\Lambda = Z^{eff}_S / Z^{eff}_L$ is of interest. Changing the interaction between the pure components (*via* $n$, $c_B$ and $Z_{eff}$) and the mixture parameters ($\Gamma$, $\Lambda$ and $p$), different situations can be obtained.

Hard sphere – polymer mixtures show entropically driven demixing of the components [15, 16] which gives rise to gas, liquid and crystal phases as well as to metastable gels, clusters and attractive glasses [17, 18, 19, 20, 21]. Also in pure hard sphere mixtures entropy plays an important role, but in addition, packing constraints dominate the phase behavior at large packing fraction. At size ratios between 0.4 and 0.8 they develop stoichiometric compounds with peritectics at certain size ratios or compositions. The predicted and experimentally confirmed structures include AB, $AlB_2$, the Laves-Phase $MgZn_2$ as well as $NaZn_{13}$ [22, 23, 24, 25, 26, 27, 28, 29, 30]. Simple eutectics, azeotropes and spindle-type phase diagrams are found only for very similar components with Γ approaching unity [25, 31]. Experimental hard sphere mixtures often tend to vitrification, induced by polydispersity and/or kinetic arrest at high density [32].

As like-charged spheres under low salt concentrations own a soft repulsion, their mixtures are less dependant on stoichiometric compositions to show a good miscibility. They are generally well miscible in both the fluid and the crystalline state. Stoichiometric compounds are only found at high densities, where the Coulomb potential is strongly screened by the counter-ions and the interaction becomes hard sphere like [33, 34, 35, 36]. Like for hard spheres, one experimentally observes a variety of phase diagram types which for increasing size ratio Γ evolve from phase separation [37], over eutectic [39, 38] and azeotropic [33, 37] behavior to spindle-type [33, 39] phase diagrams. Thus decreasing the asymmetry leads to an increased miscibility of the components in both fluid and solid [33, 39, 37]. For a case of highly charged small spheres mixed with low charged large particles, an increased miscibility in the solid phase above that of the fluid was found. The resulting phase diagram type is an upper azeotrope [40]. The general role of the charge asymmetry, however, is not yet fully understood [37]. Finally, for oppositely charged spheres a wealth of compound structures is observed [41, 42]

In a recent paper we have presented a general overview on different types of phase diagrams obtainable under variation of the size and charge ratio in binary charged sphere mixtures [37].

In the present contribution we present a detailed investigation on an exhaustively deionized charged mixture with very similar components. This mixture had previously been studied concerning its conductance, its elasticity and its nucleation kinetics [43]. From these result it was expected to show a spindle-type phase diagram. Going beyond this and other authors work [33, 39], we here show that, in fact, a broad coexistence region opens between the freezing and melting line at intermediate compositions. Moreover, the two phase boundary shows opposite curvature, thus clearly discriminating our spindle-type system from a weakly pronounced azeotrope. Finally, for the first time we attempt a comparison of the shape and location of the phase boundaries to theoretical results.

In the following we first describe the system and its conditioning and shortly introduce the experimental techniques and necessary theoretical background. We then present our results and compare the observed phase behavior to other experimental findings on charged and hard spheres as well as to theoretical expectations. We end with a short discussion on open points and future challenges.

2. Experimental

Particles

The suspensions used for the binary mixture were species PS90 and PS100B of charged colloidal polystyrene spheres in water stabilized by carboxyl surface groups purchased at Bangs Laboratories Inc. (Fishers, Indiana, USA).

The pure components were carefully characterized beforehand. As seen from Tab. I both pure components are very similar in size, charge and phase behavior. Radii refer to transmission electron microscopy performed by the manufacturer. Charge determination in colloidal systems is still an issue under discussion, as different experiments and theoretical approaches yield different numbers. For a comparison and discussion of effective charges in general, their emergence from the ideally dilute reference case and the consequences for the system properties see e.g. 44, 45, 46 and 47. In Tab. I we report the number of ionizable surface groups $N$, which was determined by conductometric titration with NaOH. The effective charge $Z^{eff}_\sigma$ from density dependent measurements of the conductivity measures the number of mobile counter-ions. It was found to be independent of the particle number density in the range of $n$ = 0.5 - 40μm$^{-3}$, which correspond to volume fractions around $\Phi = (4\pi/3)a^3 n$ = 0.003 - 0.02 [48]. The effective charge from elasticity $Z^{eff}_G$, gives a measure of the interaction strength. Also this quantity was observed to be independent of $n$ in the crystalline phase [49]. The phase behavior of the pure species was determined under exhaustively deionized conditions as a function of increasing density $n$. Crystals are conveniently detected by visual inspection of the samples. They appear as brightly iridescent objects due to their Bragg scattering in the range of visible light. The crystal structure was determined from static light scattering to be body centered cubic (bcc). The freezing points given in Tab. I are inferred from an interpolation between the densities of the last sample showing no crystals and the first sample doing so. The uncertainty given reflects this procedure. The melting points are located between the density of the last sample showing still some fluid and the first fully crystalline sample. We note that the determination of the melting points by visual inspection of samples in cylindrical vials slightly underestimates the melting density $n_M$, as small pockets of remaining melt may escape observation. The more accurate detection using Bragg microscopy in flat rectangular cells in a dilution series employing a continuous deionization procedure [50] was not performed here, as this method needs considerably larger amounts of particles and is less suitable for exhaustive deionization. Therefore, the melting density given in Tab. I represents a lower boundary. To summarize, from Tab. I we obtain a size ratio of $\Gamma$ = 0.9 and an effective charge ratio $\Lambda$ = 0.96 and very similar freezing and melting densities for the two species.

**Table I:** Properties of the pure components under deionized conditions. The source gives the manufacturer and the batch number. $2a_{nom}$ refers to the diameter determined by transmission electron microscopy with the standard derivation as error. $N$ is the titrated number of carboxylic surface groups, $Z^{eff}_\sigma$ and $Z^{eff}_G$ denote the effective charges from conductivity measurements and shear-modulus measurements, respectively. The freezing and melting number densities, $n_F$ and $n_M$, and the corresponding volume fractions $\Phi$ were obtained after batch preparation in the exhaustively deionized state.

| Suspension | Source | $2a_{nom}$/nm | N | $Z^{eff}_\sigma$ | $Z^{eff}_G$ | $n_F$/µm$^{-3}$ | $n_M$/µm$^{-3}$ | $\Phi_F$/% | $\Phi_M$/% |
|---|---|---|---|---|---|---|---|---|---|
| PS90 | Bangs, 3012 | 90±5 | 16400±100 | 504±35 | 315±8 | 1.1±0.2 | 2.0±0.5 | 0.042 | 0.076 |
| PS100B | Bangs, 3067 | 100±5 | 49900±500 | 530±38 | 327±10 | 0.6±0.2 | 1.0±0.2 | 0.031 | 0.052 |

Sample conditioning

Samples were prepared from pre-cleaned and filtered stock suspensions following an improved protocol of [33], described in detail elsewhere [38]. In short, samples of desired density $n$ (as controlled *via* static light scattering) and composition $p$ were filled into 2ml cylindrical cells (Supelco, Bellefonte, PA, USA) containing about 0.2ml mixed bed ion exchange resin (IEX, Amberlite, Carl Roth GmbH+Co.KG, Karlsruhe, Germany ) and sealed with a Teflon® septum screw cap. The samples were kept cap-down over several weeks under regular gentle stirring. To monitor the deionization process the size of the crystallites, reappearing after stirring the sample, was measured. Deionization was considered completed when this quantity had reached a constant low value after some two months [51].

The residual ion concentration depends on the quality of the IEX and the deionization protocol followed. For the batch procedure described it is not measurable *in-situ*. From the comparison of the obtained freezing densities (c.f. Tab. I) to those measured earlier in a closed tubing system at $c_S$ = 0.2µmol/l [50] we infer a considerably improved degree of deionization in our case. The theoretical limit of the residual ion concentration is given by the self-dissociation of water: [H+][OH-] = $1 \times 10^{-14}$ (mol/l)$^2$. We identify the proton concentration with the amount of counter-ions released by the particles: [H+] = $n\, Z^{eff}_\sigma / (1000\, N_A)$, where $Z^{eff}_\sigma$ is the effective charge of the particles determined by conductivity measurements [52] and $N_A$ the Avogadro's number. Then, for typical cases investigated here the minimal residual ion concentration is estimated to be on the order of $c_s \cong$ [OH-] $\leq 1 \times 10^{-8}$ mol/l.

Interaction

The experimental parameters varied in our experiments are the total number density $n = n_S + n_L$ and the composition $p = n_S / (n_S + n_L)$ which is given as the number ratio of the components. Their variation is reflected in a variation of the pair interaction energy $V(\bar{d})$, where $\bar{d} = 1/n^{1/3}$ is the average particle distance. For isotropic binary mixtures Lindsay and Chaikin [53] suggested to construct a suitable $V(\bar{d})$ accounting for the $p$-dependent average particle charge and the corresponding screening parameter:

$$V(\bar{d}) = \frac{1}{4\pi\varepsilon\varepsilon_0}(p^2\tilde{Z}_S^2 + 2p(1-p)\tilde{Z}_S\tilde{Z}_L + (1-p)^2\tilde{Z}_S^2)\frac{e^{-\kappa\bar{d}}}{\bar{d}} \qquad (1)$$

with

$$\tilde{Z}_{S,L} = Z^{eff}_{G,S,L}\frac{e^{\kappa a_{S,L}}}{1+\kappa a_{S,L}} \qquad (2)$$

and the inverse screening length

$$\kappa = \sqrt{\frac{e^2}{\varepsilon\varepsilon_0 k_B T}(1000 N_A c_s + pnZ^{eff}_{G,S} + (1-p)nZ^{eff}_{G,L})} \,. \qquad (3)$$

Here $\varepsilon\varepsilon_0$ is the dielectric permittivity of the suspending medium, $e$ is the elementary charge and $k_BT$ is the thermal energy. Note that the use of $\bar{d}$, which is defined for isotropic systems, neglects the small difference to the nearest neighbor distance in the crystal phase [54]. Note further, that as suggested by [53], the effective charge from elasticity measurements $Z^{eff}_G$ is used. The effective elasticity charge reflects the influence of triplet and higher order interactions [44, 55]. In contrast to the larger effective charge derived from conductivity (describing the number of uncondensed counter-ions), this kind of effective charge gives a very accurate description of the phase behavior of single component systems [11]. In addition, we carefully checked that the chosen interaction energy, Eqn. (1), describes the small linear changes in elasticity and solidification kinetics with composition very well [43].

To facilitate comparison to other systems, we will present our results on the phase behavior in different representations. The first is in terms of the experimental parameters $n$ and $p$. The second was suggested by Meller and Stavans [39] in their investigation on strongly screened charged spheres. Here the inverse of the total volume fraction $\Phi^{-1} = (\Phi_S + \Phi_L)^{-1}$ is plotted versus the composition. As a representation of the interaction strength of the system instead of only the density, the phase diagram may be given in terms of an effective temperature $T^{eff} = k_BT / V(\bar{d})$ over composition defined as ratio between thermal and pair interaction energy at the nearest neighbor distance $V(\bar{d})$. Note though, that through the density dependence of $V(\bar{d})$ the effective temperature is still a density variable. Therefore, even for the pure components a finite coexistence region will appear in the $T^{eff}$-direction of this phase diagram.

In addition we will also show universal phase diagrams originally suggested by [4] for single component Yukawa particles and often used in theoretical studies [5, 6, 7, 8, 9]. The first will be in terms of the effective temperature $T^{eff}$ and a coupling parameter $\lambda = \kappa\bar{d}$. The second is in terms of the Einstein temperature $\tilde{T} = k_B T / m\omega_E^2 \bar{d}^2$, relating to the typical oscillation frequency $\omega_E$ of a particle in a fcc or bcc structure including the sum of the interaction energies up to the third next neighbors in a crystalline environment [4, 8]. The coupling in this second representation is expressed in terms of $\kappa a_{WS}$, where $a_{WS} = (4\pi/3)^{-1/3} \bar{d}$ is the radius of the spherical Wigner Seitz cell. The effective temperature $\tilde{T}$ of the experimental data is calculated using the normalized total interaction energy given in [4], fitted by a sum of three exponential decays.

3. Results

In Fig. 1 we show a exemplary Debye-Scherrer patterns of our mixture recorded at a particle density of $n = (3.9 - 6)\mu m^{-3}$ ($\Phi = 0.0016 - 0.0033$) as indicated and compositions of $p = 0, 0.2, 0.4, 0.6, 0.8, 1$. Shown data are representative for all other investigated compositions and densities. Depending on particle number density up to four Bragg peaks are accessible. The Miller indices corresponding to the peaks are indicated and the peak positions follow $q_{hkl} = (2\pi/g)* (h^2+k^2+l^2)^{1/2}$, where g is the lattice constant of the cubic unit cell. This allows identifying the crystal structure to be body centered cubic. Moreover, the absence of additional superstructure peaks at low scattering vector suggests that substitutional alloys with no compositional order are formed. If compounds were formed, their additional reflections should be well visible even at the present large size ratio [33, 34]. Our observations agree favorably with earlier studies reported for this and other large $\Gamma$ systems [56].

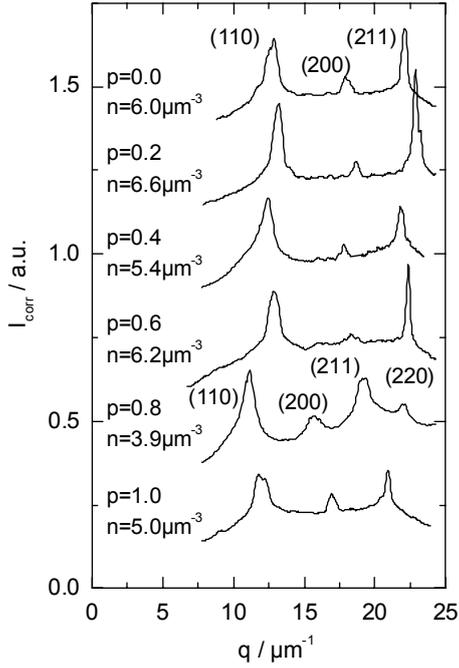

**Fig. 1**: Representative Debye-Scherrer pattern for samples with indicated composition and particle number density. Up to four diffraction peaks are clearly discernible and are Miller indexed. The corresponding structure is body centered cubic for both the pure components and the mixtures.

In our determination of the phase diagram we are therefore only concerned with a simple fluid-solid transition. Different representations of the experimentally observed phase behavior are shown in Fig. 2 a - c. In Fig. 2a we plot the raw data in terms of the number densities $n$ and the composition $p$. Symbols mark the phase states observed a few days after last shaking. Black squares refer to the bcc crystalline phase, the cyan diamonds denote fluid/crystalline coexistence and the red open circles indicate fluid phase. The red and black solid lines are a guide to the eye to indicate the position of the freezing line (liquidus) and melting line (solidus). To construct these lines we took for each $p$ the midpoints in $n$-direction between adjacent points of different phases and placed a spline through these data. The pure components show a finite coexistence region at roughly similar densities. With increasing $p$ the melting line bends downward then upward again. At the same time the freezing line first beds upward then downward again. The resulting coexistence region shape is a thick, approximately horizontally oriented spindle.

Fig. 2b shows the same data and lines in the $\Phi^{-1} - p$ representation as originally suggested by [39]. Shape and orientation of the coexistence region are retained. Fig. 2c shows the $T^{eff} - p$ representation accounting for the combined influence of density and screening on the interaction energy. For the investigated number densities, $T^{eff}$ is a monotone function. Again a horizontally oriented spindle is observed. Note that here, however, the data points for the crystal phase do not show a strong variation in $T^{eff}$ with density and all nearly coincide. The melting line (solidus) here appears much better defined than before and in addition is markedly flattened. As compared to previous work showing a single phase boundary [33, 39], we here were able to determine both liquidus and solidus.

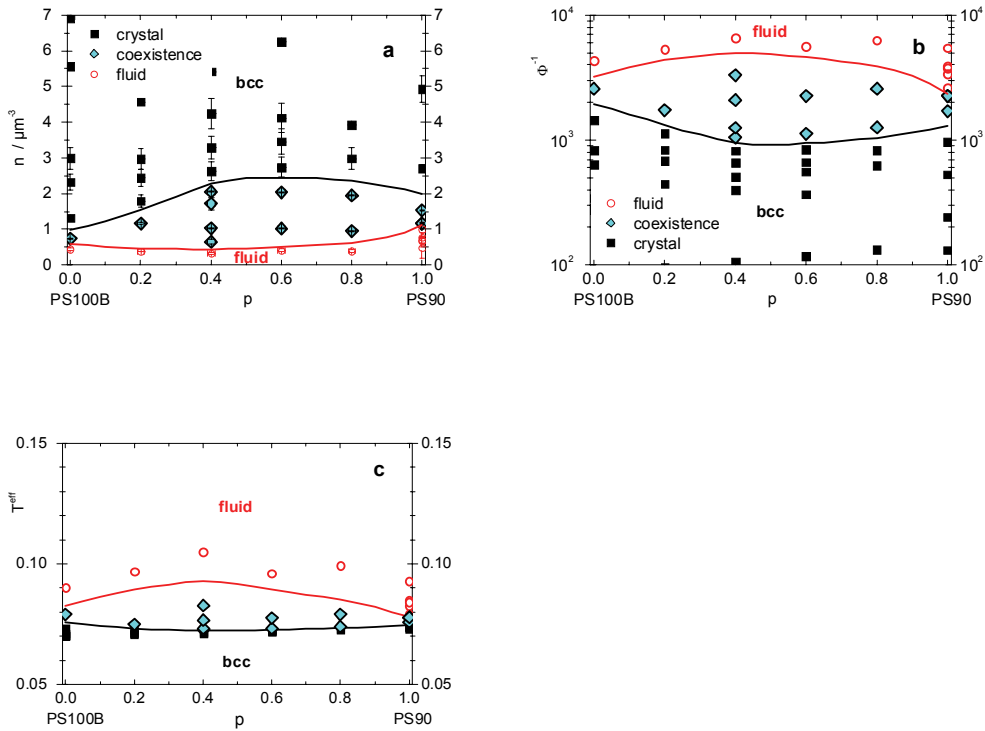

**Fig. 2**: (color online) Binary phase diagrams of the mixture PS90/PS100B in different representations. Symbols denote fluid phase (open red circles), crystal phase (black filled squares) and the coexistence region (cyan diamonds). Liquidus (red) and solidus (black) as determined in the *n – p* representation and transferred to the other representations are given by the solid lines. a) Density – composition plane; the uncertainty in density is given by the error bars, while for the composition it is of the order of ±0.02. b) Semi-log plot of the inverse volume fraction – composition plane; c) effective temperature – composition plane.

4. Discussion

Three points deserve our particular attention in this discussion: first, the shape and type of phase diagram, second the broadening of the coexistence region and third the location of the phase transition in comparison to theoretical expectations.

The melting densities of the pure components of our binary charged sphere mixture with size ratio $\Gamma$ = 0.9 and effective charge ratio $\Lambda$ = 0.96 were very similar. To unequivocally discriminate between spindle-type and azeotropic phase behavior, we demonstrated the opposite behavior of the melting and freezing lines as a function of composition. The spindle-type phase diagram is retained in all different representations. Generally, spindle-type phase diagrams are expected for small differences in the particle properties and an equally excellent miscibility in both fluid and solid. Previous simulations of hard spheres at $\Gamma$ = 0.97 [31] as well as measurements of charged spheres at $\Gamma$ = 0.87 and elevated salt concentrations [39] and at $\Gamma$ = 0.76 under deionized conditions [33] showed spindle-type phase diagrams. There, however, a monotonic evolution of the transition density between those of the pure components was observed, as these were more dissimilar than in the present study. Our study also confirms the trend that under deionized conditions the range of the Coulomb repulsion for both sphere species is increased and in our case the spindle is located at much smaller freezing densities than in previous studies on salty systems or hard spheres.

In addition to the location and orientation of the spindle we here also could determine the extension of the coexistence region, which shows some interesting features. For the pure components the relative density jump amounts to $\Delta n = (n_M - n_F) / n_M \cong 0.4$. The maximum extension of the coexistence region at $p = 0.55$ is $(n_M - n_F) \approx 2\mu m^{-3}$. This yields $\Delta n \approx 0.8$. The width in $T^{eff}$ direction for the pure components is $\Delta T^{eff} = (T^{eff}_F - T^{eff}_M) / T^{eff}_F = 0.09$ and increases to $\Delta T^{eff} = 0.31$ at $p = 0.55$.

The width of the coexistence region may be influenced by several effects, including interaction kind and strength and polydispersity. For monodisperse hard spheres, the density jump at coexistence is $\Delta n = 0.09$. Monte Carlo simulations on binary mixtures of monodisperse hard spheres at $\Gamma = 0.97$ showed that the fluid phase was slightly enriched in small particles [31]. Unfortunately the coexisting densities were not reported. For single component hard spheres of increasing polydispersity $\Delta n$ was found to decrease continuously. As the fluid phase tolerates a higher polydispersity than the solid phase, the number density difference becomes negative at high polydispersities while, at the same time, the coexisting solid has a higher volume fraction. I. e. also here the large particles segregate into the solid [14].

Monodisperse charged sphere systems are composed of macro-ions, counter-ions and other micro-ions stemming from added salt. Density functional calculations which explicitly accounted for the micro-ionic degrees of freedom have been performed by Graf and Löwen [13]. The authors found lower $\Delta n$ than for hard spheres and observed the fluid phase to be enriched with micro-ions. They further reported a decrease of $\Delta n$ with decreasing salt concentration and at salt concentrations below $1\mu mol\ l^{-1}$ values of $\Delta n = 0.02 - 0.03$ were found. Similarly, Hynninen and Dijkstra [57] report from computer simulations, that the gap width reduces to zero upon charging up ensembles of monodisperse hard spheres. No studies on charged sphere mixtures or on the influence of polydispersity are as yet available, partly because theoretically the effect is expected to be small.

On the experimental side only few data are available on the width of coexistence regions. For single component systems similar widths as reported here for $p = 1$ and $p = 0$ were found by [1] at similar concentrations of deionized particles, and by [2] for high particle concentrations and millimolar salt concentrations. There also a wide bcc-fcc coexistence was reported, which was later also confirmed for deionized systems by [12]. In collaboration with others, the present authors recently reported the phase diagrams of a large number of binary charged sphere mixtures of different size and charge ratio [37]. For an azeotropic mixture with $\Gamma=0.82$ the maximum width of the coexistence region, found slightly off the azeotropic point, was on the same order of magnitude than reported here. For a eutectic mixture of $\Gamma=0.57$ under the influence of gravity it was found to be even larger around the eutectic point [38].

Our data may further be compared to the liquidus-solidus separation for metal alloys in a temperature – composition representation: $\Delta T = (T_L - T_S)/T_L$, where $L$ and $S$ refer to the liquidus and solidus temperature, respectively. For instance for Ni-Cu the maximum extension is some 60K at a liquidus temperature of about 1500K, yielding $\Delta T = 0.04$ [58].

As compared to theoretical expectations and data on metals but in line with observations on more dissimilar charge sphere mixtures we here observe a remarkably pronounced widening of the coexistence region for intermediate compositions. In fact, if we consider that the values obtained for $n_M$ are only lower bounds, the true widening may be even broader. At present, the underlying reason for this observation is not understood.

We therefore feel free to indicate some speculative lines of reasoning. First, the pair-wise additive, effective one component hard-core Yukawa description may fail for charged sphere mixtures in several points. The issue of many body interactions is already accounted for by the use of the effective elasticity charge. For the general location of the melting line, an effective (monodisperse) one component description seems to work reasonably well (as will be shown below in the comparison to theoretical expectations). This may, however, not be sufficient for the description of the broadening of the coexistence region. In particular, no depletion effects are included in this description, which at very small size ratios lead to complete phase separation [37]. It is thinkable, that such effects could lead to a partial segregation at larger size ratios. Very recently, combined packing constraints and depletion effects have been observed to lead to strongly fractionated crystallization in systems of spheres interacting *via* a $r^{-12}$ potential [59]. The present potentials are less steep, but, if we map our system onto an effective one component hard sphere system [60], the combined polydispersity is on the order of 10%. This is way above the value of 5% found for the onset of segregation in hard spheres [14] and would be in line with the experimentally observed increase in the gap width with decreasing size and charge ratio [37]. Second, our analysis only considered the differences in *n* but – as these were not measurable – neglected possible differences in the salt concentration. Third, our experiments were carried out under gravity, such that the crystalline phase always resided at the cell bottom. Combining both points, the work of van Roij [61] and in particular of Torres et al. [62, 63] would suggest, that a gravity driven segregation could stratify the system into two layers, each enriched with one of the species. To balance the osmotic pressures and chemical potentials, the salt concentrations would have to adjust, too. Summarizing, we would not be too surprised, if the observed large gap width would connected to some segregation effect. Segregation was experimentally confirmed *via* radius measurements on diluted samples taken of two coexisting phases in a mixture of size ratio $\Gamma = 0.57$[38]. This could be possible also here, although the considerably smaller size ratio definitely presents a challenge.

The third point of our discussion concerns the experimental phase diagrams in comparison with theoretical expectations based on an effective one-component description. The pair energy $V(\bar{d})$ suggested by Lindsay and Chaikin (Eqns. 1 - 3) was derived to describe the energetics in compositionally disordered solid solutions. It has been successfully used to describe the composition dependence of the elasticity and other properties in deionized charged sphere mixtures including the present one [43]. Therefore it is an interesting question, how well it performs in describing the location and shape of the phase boundaries observed for our mixture. In contrast to the situation for hard spheres no theoretical expectations for the phase behavior of charged sphere mixtures are as yet available. Only for single component charged spheres predictions of the phase behavior have been given [4, 5, 6, 7, 8, 9]. Lacking more advanced treatments we first compare our data to the prediction of [4] in the $T^{eff}$ - $\lambda$ representation of Fig. 3a,b, then to the other theoretical predictions in the $\tilde{T}$ - $\kappa a_{WS}$ representation [4, 5, 6, 7, 8] in Fig. 4a,b. As discussed above, the effective temperature in the former corresponds to the pair energy at the nearest neighbor distance, while in the latter it corresponds to the total interaction energy per particle up to the 3$^{rd}$ nearest neighbors. All quantities were calculated using the particle properties given in Table I, the ion concentration at the theoretical minimum and employing Eqns. (1) - (3) [53].

In both Figs. the filled diamonds representing the melting points of the pure samples are located on the so-called state lines (thin solid lines) [11]. These represent the states of our samples in the $T^{eff}$ - $\lambda$ and $\tilde{T}$ - $\kappa a_{WS}$ planes, which are attained, when at fixed composition, effective charge and salt concentration the number density is increased. In both drawings they show the known characteristic behavior. Starting from low particle concentrations at the top of the $T^{eff}$ - $\lambda$ representation, the effective temperature first decreases with increased $n$ while the coupling parameter $\lambda = \kappa \bar{d}$ decreases due to the decreased average particle distance at constant $\kappa$. When the particle concentration becomes large enough to increase the screening parameter through the increased concentration of counter-ions, the curves turn right. At the same time the decrease of the effective temperature is slowed or even reversed, due to the increased screening. Only at much larger particle densities the effective temperature decreases again strongly, as now the hard sphere limit is approached. These prominent features nearly vanish in the $\tilde{T}$ - $\kappa a_{WS}$ representation, as here the curves are straightened due to the division by $\lambda^2$. The resulting state lines are much closer to each other and proceeding monotonously and much more steeply.

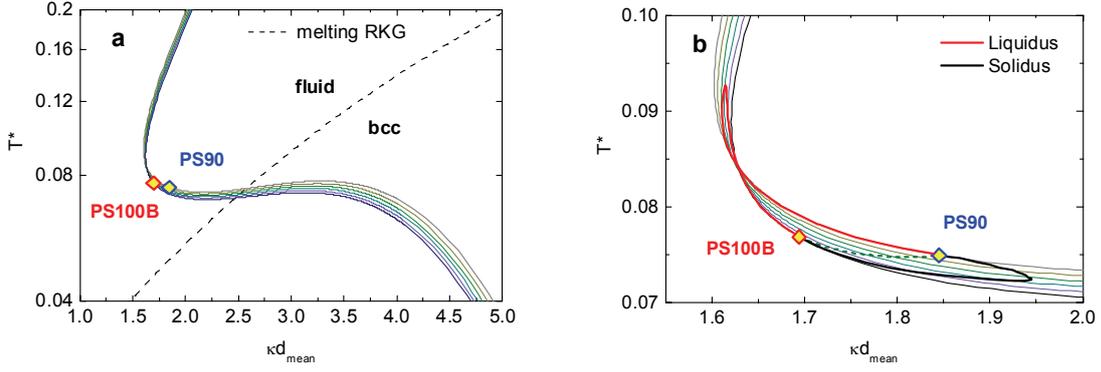

**Fig. 3:** (color online) Experimental melting points in the $T^{eff}$ - $\lambda$ representation. The thin colored lines represent the state lines of the investigated systems calculated using the effective charge from Tab. I, a minimum salt ion concentration and a particle density, which increases from top left to bottom right. The composition $p$ was varied in steps of 0.2. a) The melting points (diamonds) of the exhaustively deionized pure components PS90 (blue) and PS100b (red) are located slightly above the universal melting line obtained for pair-wise additive Yukawa interactions [4]. b) Close-up of the transition region. The thick upper red and lower black lines represent the freezing and melting curves shown in Fig. 1a. The thick, dashed olive line is a guide to the eye showing linear interpolation between the pure component melting points in Fig. 1a.

The two diamonds indicate the location of the pure component melting points. In Fig. 3a they are located slightly above and to the left of the universal melting line of Robbins et al. ([4], thin, black dashed line). This observation is in line with those made previously for this and other one component systems at slightly larger salt concentrations ($c_B = 2 \times 10^{-7}$ mol l$^{-1}$) [11]. Similar behavior can be seen in Fig. 4a, where we show the phase diagram in the $\tilde{T}$ - $\kappa a_{WS}$ representation. Various theoretical predictions of [4] (thin dashed line), Hamaguchi et al. ([8], red open symbols, thin solid line) and numerous other simulation results collected in [8] (open symbols [5], [7] and [6]) enable a precise comparison with the experimental data. The compilation of the different results gives a measure of the discrepancies arising from the different methods and accuracy of the simulation data. The experimentally observed crystal phase is somewhat more stable than the crystal phase observed in the simulations.

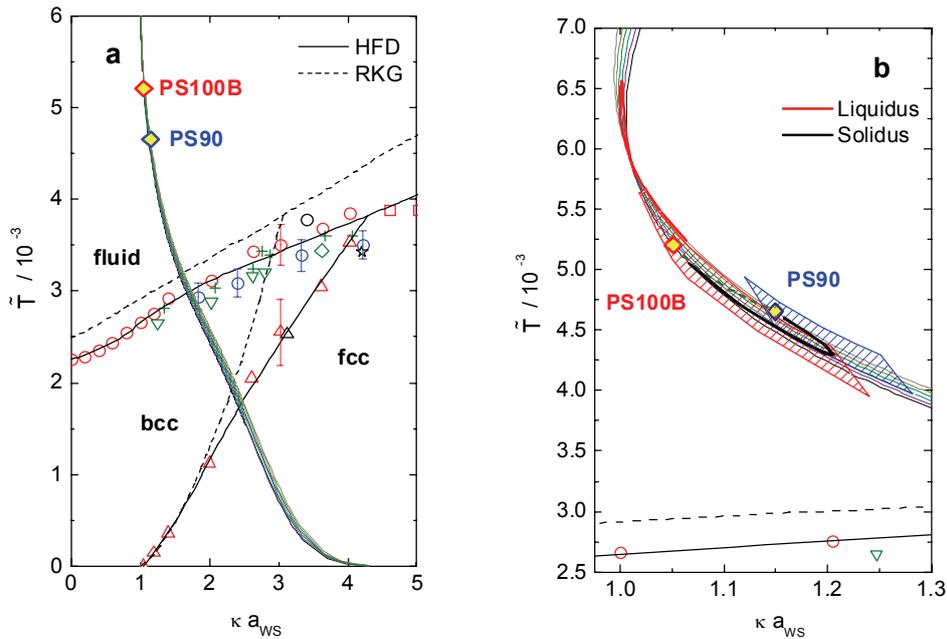

**Fig. 4:** (Color online) Experimental melting points and state lines (colored, solid lines) of PS90 (blue diamond) and PS100b (red diamond) in the $\tilde{T}$ - $\kappa a_{WS}$ representation. a) Comparison to several theoretical predictions of the phase behavior. We show the fitted phase transition lines given by Hamaguchi et al. (HFD, thin black lines, [8]) and Robbins et al. (RKG, thin dashed black lines, [4]). We further give the locations of the melting points (circles) and of the bcc-fcc transitions (up triangles) as obtained by Meijer and Frenkel (blue,

[5]), Du Pont et al. (black, [7]) and HFD (red, [8]). The triple point of Du Pont et al. is given by the star; the location of fluid (plus) bcc (down triangle) and fcc (diamond) structure as obtained by Stevens and Robbins [6] are given by olive symbols. b) Close-up of the transition region with the melting points of the two pure species surrounded by error regions (hatched) accounting for the statistical uncertainties in $Z^{eff}$, $n$, as well as the systematic uncertainty in $n_M$.

In Fig. 3b and 4b we show close-ups of the respective transition regions. Now also the phase diagram shape is drawn. The dashed olive line is a guide to the eye, corresponding to a linear interpolation between the pure component melting points in Fig. 2a, while the solid lines represent the actually measured freezing (red) and melting (black) lines. In Fig. 4b we additionally show the experimental error regions resulting from the small statistical uncertainties in the effective charge $Z^{eff}_\sigma$ entering the salt concentration estimate as well as $Z^{eff}_G$ and the number density $n$ of the individual samples entering the calculation of the melting curve shape. In addition we account for the much larger systematic uncertainty in locating the phase transition points in Fig. 1. This does not affect the melting curve shape. Rather, it influences the exact location of the melting curve. Concerning the former, we see that the phase diagram shape deviates qualitatively from the spindle-type seen in Fig. 1. Due to the fact that the state lines for the pure components cross slightly above their effective melting temperatures, the freezing line performs a loop. On the other side, the shape of the melting line is retained, as in the regions of interest no state line crossing occurs. Quite expectedly, this shape is qualitatively incompatible with the nearly straight course of the universal melting line predicted for a single component system. Concerning the latter, we see a quantitative deviation of the experimental melting curve from the universal melting curve in both figures. This deviation, however, is not larger than the deviation seen for the pure samples.

5.  Conclusion

We have studied the phase behavior of a binary charged sphere mixture of two species similar in size and charge under conditions of very long ranged repulsions. In our case the melting and freezing densities of the pure components were both rather similar. We could, however, demonstrate that in terms of composition, the melting and freezing lines show opposite trends and therefore unequivocally determine the type of phase diagram as spindle-like. Similar to systems of other interactions, like metals or hard spheres, the crystalline phase was of body centered cubic structure with no compositional order, i.e. a substitutional alloy. We therefore may conclude that our two species of charged spheres show an indifferent miscibility, i.e. they are highly and equally well miscible in both phases.

The strength and long-ranged nature of the electrostatic interactions between our highly charged spheres in thoroughly deionized suspension gave rise to transition densities on the order of one particle per cubic micron. The comparison with results from several theoretical studies in $T^{eff}$ - $\lambda$ and $\tilde{T}$ - $\kappa a_{WS}$ representations show no qualitative description of the solidus but a quantitative comparability with the experimental data similar to that of the pure components. We therefore infer that the employed potential gives a decent description of the interactions within our highly miscible mixture. *Vice versa,* the theoretical predictions for single component Yukawa systems may provide a reasonably accurate first estimate for the location of the melting line in experimental charged sphere mixtures with size and charge ratios close to unity.

Finally, our results also revealed a large density jump across coexistence amounting to $\Delta n \approx$ 0.8 at $p = 0.55$, which is neither expected from previous experimental and theoretical work on single component systems, nor from simulations on binary mixtures of hard spheres. Unfortunately, a general prediction of the rich and complex phase behavior of charged sphere mixtures is not yet available. A study of the width of coexistence regions for highly miscible charged sphere mixtures including the possibility of differing compositions for fluid and solid phase may, however, be a rewarding starting point.

Acknowledgements

The authors are pleased to thank Patrick Wette, Hans Joachim Schöpe, Dieter Herlach and Hartmut Löwen for numerous fruitful discussions. Financial support by the DFG (Pa459/14 and Pa459/16) as well as by the former Material Science Research Centre (MWFZ), Mainz is gratefully acknowledged.